# Quantifying the Impact of Large Language Models on Collective Opinion Dynamics

**Chao Li[a], Xing Su[a]*, Haoying Han[a], Cong Xue[a], Chunmo Zheng[a], Chao Fan[b]**

a College of Civil Engineering and Architecture, Zhejiang University, Hangzhou, Zhejiang, 310000
b College of Engineering, Computing, and Applied Sciences, Clemson University, Clemson, SC, 29631.

**Abstract**

The process of opinion expression and exchange is a critical component of democratic societies. As people interact with large language models (LLMs) in the opinion shaping process different from traditional media, the impacts of LLMs are increasingly recognized and being concerned. However, the knowledge about how LLMs affect the process of opinion expression and exchange of social opinion networks is very limited. Here, we create an opinion network dynamics model to encode the opinions of LLMs, cognitive acceptability and usage strategies of individuals, and simulate the impact of LLMs on opinion dynamics in a variety of scenarios. The outcomes of the simulations inform about effective demand-oriented opinion network interventions. The results from this study suggested that the output opinion of LLMs has a unique and positive effect on the collective opinion difference. The marginal effect of cognitive acceptability on collective opinion formation is nonlinear and shows a decreasing trend. When people partially rely on LLMs, the exchange process of opinion becomes more intense and the diversity of opinion becomes more favorable. In fact, there is 38.6% more opinion diversity when people all partially rely on LLMs, compared to prohibiting the use of LLMs entirely. The optimal diversity of opinion was found when the fractions of people who do not use, partially rely on, and fully rely on LLMs reached roughly 4:12:1. Our experiments also find that introducing extra agents with opposite/neutral/random opinions, we can effectively mitigate the impact of biased/toxic output from LLMs. Our findings provide valuable insights into opinion dynamics in the age of LLMs, highlighting the need for customized interventions tailored to specific scenarios to address the drawbacks of improper output and use of LLMs.

## Introduction

The process of opinion expression and exchange can facilitate the interactions of diverse perspectives and enables individuals to make informed decisions and participate in civic life[1-3]. It has been argued that social interactions, such as face-to-face and through traditional media (e.g. TV, newspapers, Twitter), are fundamental in this process[4-7]. This process thus has been extensively studied in the past decades, with several opinion models proposed and modified in the context of traditional media[8-13]. However, despite the advances and evolution of these opinion models, such as considering agent stubbornness[14,15] and noise[16], they still fail to fully capture the impacts of LLMs on collective opinion dynamics.

Traditional media mainly rely on widespread information dissemination such as radio or broadcast television[17,18], or directly bridge the gap between individuals such as Twitter[19-21], to influence collective opinion networks. Specifically, traditional media opinions are manually reviewed and validated before output, so the output is more trustworthy, unbiased and accurate[22]. The opinion delivery process of traditional media is a one-way direct interaction, i.e., they influence the public through the unilateral information dissemination[23]. Unlike the pattern of traditional media, LLMs play the role of a personal copilot to affect collective opinion networks through their penetration of personal opinions. Fig.1 shows that there are significant differences between LLMs and traditional media in terms of opinion shaping process, opinion interaction and opinion output. LLMs will only be manually reviewed during the opinion formation process[24]. Due to documented limitations in resources and problematic patterns in training data, it is highly possible to contain false, biased and toxic content[25-27]. Hence, the output of LLMs will carry these contents, and the opinion delivery process will be a two-way interaction. That is, LLMs influence individuals through a question and answer (Q&A) format of interaction, a pattern that disseminates the output of LLMs more efficiently. Meanwhile, as LLMs become more prevalent in our daily lives (such as ChatGPT[28], the fastest-growing consumer app ever, with hundreds of millions of active users just two months after launch[29]), the risks shown in Fig.1 have been recognized as an urgent problem, leading to the emergence of many other problems such as leaking private information[30] and overreliance[31,32]. Different organizations and individuals introduce different usage strategies, even many of them choose to neglect the benefits of LLMs and completely prohibit these effective tools for aforementioned issues.

*Corresponding author: Xing Su. Email address: xsu@zju.edu.cn.

Given the aforementioned differences and existing usage problems, it is indispensable to understand how the output of LLMs affects collective opinion dynamics, and what usage strategies should be tailored to address the drawbacks of LLMs in specific scenarios. Attempts to study the impact of LLMs on opinion have recently emerged[25,33]. Among these studies, the impacts of cognitive bias hidden in the output of LLMs have gained significant attention[34-37]. These biases include gender[38], race[39], nationality[40], and even political topics[41]. Most of this research focuses on establishing robust evidence of the existence of bias by conducting control experiments with and without specific treatment. Such studies typically only consider the impacts of specific cognitive biases induced by LLMs on individual opinions, but neglect the various use strategies for LLMs that influence the opinion evolution over time.

To address these limitations, we propose a new opinion model, based on the classic Hegselmann-Krause (HK) model[42] and incorporate the bidirectional opinion shaping process and personalized usage strategies of LLMs, to investigate the dynamic evolution in opinion networks. Specifically, we categorized agents into three categories according to three different usage strategies, i.e., Nodes only Influenced by Neighbors (NIN) for no use, Nodes Influenced by Neighbors and LLMs (NINL) for partial reliance and Nodes only Influenced by LLMs (NIL) for full reliance. To mimic the reality of opinion interaction patterns of LLMs, we also propose three modifications to the HK model by taking the authoritative effect, stubbornness degree and arbitrary events of reality into account. The detailed assumptions, parameter settings and update conditions are shown in the Method section.

By implementing the proposed model, we first compared several scenarios with or without LLMs to determine if LLM has an impact on opinion dynamics. Considering its computational efficiency, we then identify parameters that have great effects on the results of the opinion dynamics, using the benchmark scenario as a reference. The detailed definitions and value ranges of original parameters are provided in Tab.1. We then perform millions of simulations to capture the evolutionary process and final formation of the opinion and explored their correlation with the filtered parameters. The detailed definitions and value ranges of indicators that are used to characterize the evolution and formation of the opinions are provided in Tab.2. Finally, we summarize the potential risks of LLMs on opinion dynamics based on the correlation matrix, explained by prior knowledge from existing studies, and investigate countermeasures for possible hazards. The results of these experiments inform us about effective usage strategies and interventions of LLMs oriented to different scenarios.

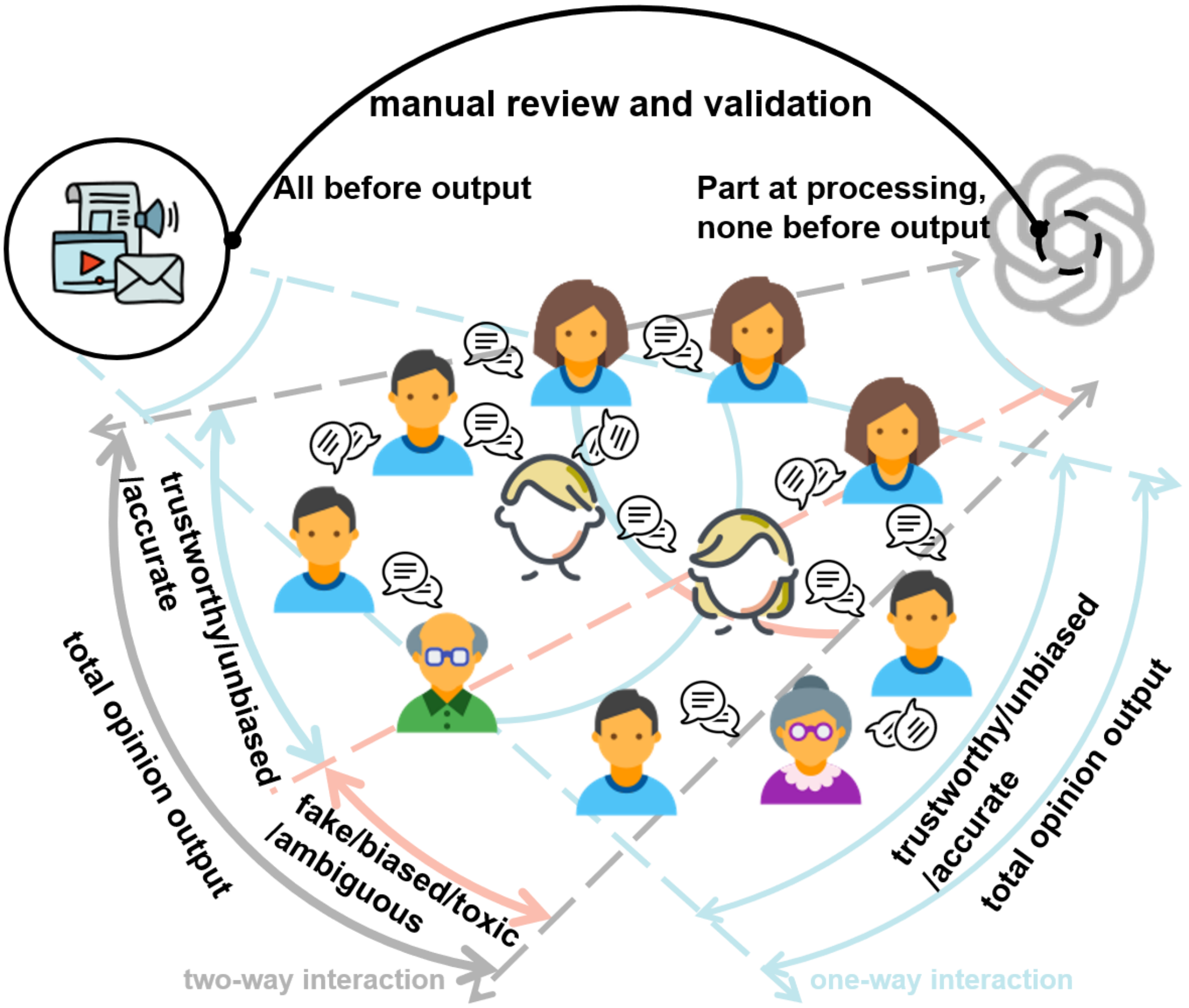

**Fig.1.** Schematic diagram of the difference between LLMs and traditional media. The left side represents the pattern of opinion dissemination in the interactions between traditional media and people, the right side represents the pattern of opinion dissemination in the interactions between LLMs and people, and the center part represents face-to-face interactions in opinion networks.

**Tab.1**. Seven controlled parameters in our modified opinion dynamic models.

| Parameter | Definition | Value range |
|---|---|---|
| $N$ | Number of group size | [0, ∞] |
| $T$ | Number of opinion exchanges | [0, ∞] |
| $\varepsilon$ | Cognitive acceptability of each agent. A value of 0 means a very low | [0, 1] |
| | acceptability of other opinions, and a value of 1 means a very high acceptability of other opinions | |
| *pro_NIN* | Proportion of the population who do not use LLMs | [0, 1] |
| *pro_NINL* | Proportion of the population who partially rely on use LLMs | [0, 1] |
| *pro_NIL* | Proportion of the population who fully rely on use LLMs | [0, 1] |
| $x_{LLM}$ | Output opinion of LLMs. A value of −1.0 means a very negative opinion on the topic, and a value of 1 means a very positive opinion on the topic | [-1, 1] |

**Tab.2**. Four indicators in our modified opinion dynamic models.

| Dimension | Indicator | Definition | Value range |
|---|---|---|---|
| Opinion evolution | $NODE_{diff}$ | Mean opinion difference of different categories of nodes. This indicator represents the evolution of the value of opinion on a topic. This indicator is minimized when all nodes have an initial opinion value of 1 and a final opinion value of -1, and is maximized when all nodes have an initial opinion value of -1 and a final opinion value of 1 | [-2, 2] |
| | $NODE_{conv}$ | Mean opinion convergence time of different categories of nodes. This indicator represents the timestep it takes for opinion to evolve to a stable state. This indicator is maximized when the exchange of opinions has not converged after the completion of the exchange of opinions, the conditions for determining convergence are shown in Eq(6) | [0, *T*] |
| Opinion formation | $NODE_{SD}$ | Mean opinion standard deviation of different categories of nodes. This indicator represents the degree of dispersion of a group's opinions relative to the mean value. This indicator is maximized when half of the nodes (n represents the number of nodes) have an opinion value of 1 and the other half have an opinion value of -1 | $[0, \sqrt{\frac{n}{n-1}}]$ |
| | $NODE_{clus}$ | Mean number of opinion clusters of different categories of nodes. This indicator represents the clustering of opinions, with a value of 1 indicating consensus and a value of 2 indicating polarization, with larger values indicating more fragmented opinions. This indicator is maximized when all opinions of nodes are inconsistent | [0, n] |

## Results

**Five crucial parameters of the model.** In Fig.2, a noticeable discrepancy between G1 and the other scenarios, specifically Fig.2A and Fig.2C, highlights the significant impact of LLMs on the opinion dynamics network. Comparing the benchmark with the remaining six scenarios, we observe that in Fig.2A, Fig.2B, and Fig.2C, the benchmark curve closely aligns with the curve of $N$=300, $T$=300, indicating an insignificant effect of an excessive number of nodes and iterations on the mean opinion value, mean opinion change value, and mean standard deviation. Fig.2B also shows that in all scenarios, opinion values almost stop changing after the number of iterations reaches 60, i.e., all individuals' opinions reach a steady state before 100 iterations are completed. In contrast, the benchmark curve differs significantly from the curves for $\varepsilon$ = 0.8, *pro_NINL* = 0.6, *pro_NIL* = 0.6, and $x_{LLM}$ = 1, demonstrating a substantial impact of modifying the threshold, the ratio of the three agents, and the output value of LLMs on the results. In Fig.2D, the benchmark curve almost entirely overlaps with the curve of $T$ = 300, indicating a minimal effect of additional iterations on the number of clusters. However, the curve of $N$ = 300 exhibits more clusters throughout the process compared to the benchmark curve, with discrepancies remaining relatively stable at around 10. Additionally, the curves of other scenarios and benchmark curves demonstrate irregular interweaving.

These findings show that the threshold, the proportion of the three agents, and the output value of LLMs significantly influence the opinion exchange process as well as the final distribution of opinions. In the rest of this paper, we will explain why and to what extent these five parameters as independent variables influence collective opinion dynamics.

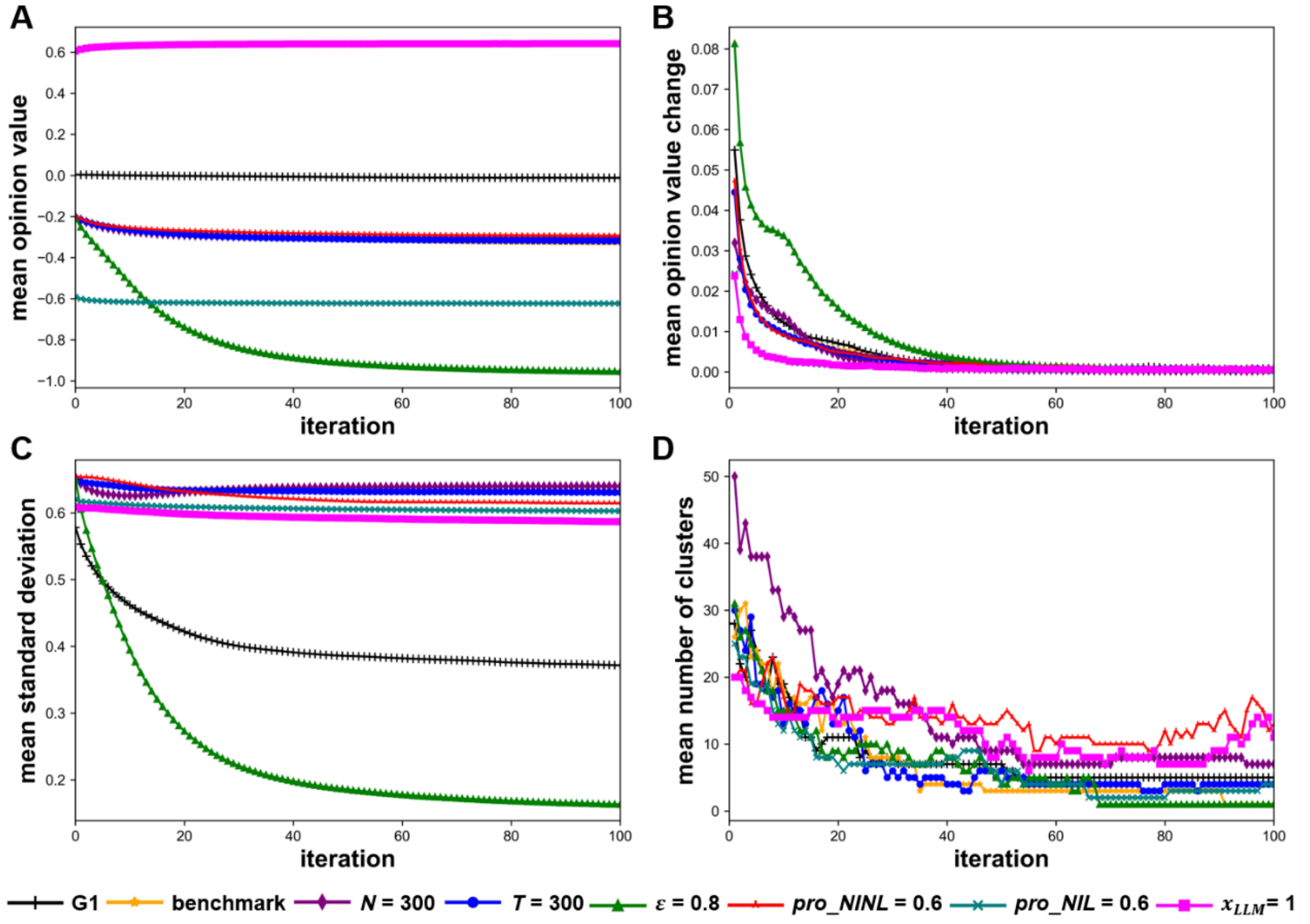

**Fig.2.** Different parameters with varying degrees of impact on the results of the opinion dynamics. For ease of exposition, the network dynamic parameters in the later will all be expressed in the order of ($N$, $T$,$\varepsilon$, *pro_NIN*, *pro_NINL*, *pro_NIL*, $x_{LLM}$). We first selected the original opinion network that is not affected by LLM (G1), then determine one of the opinion networks affected by LLM as the benchmark, and then change one of the seven

parameters at a time, so that there are a total of 8 scenarios of opinion dynamics, which are [G1(*100, 100, 0.4*); benchmark(*100, 100, 0.4, 0.6, 0.2 , 0.2, -1*); *N*=300 *(300,100, 0.4, 0.6, 0.2 , 0.2, -1*); *T*=300 *(100,300, 0.4, 0.6, 0.2 , 0.2, -1*);$\varepsilon$=0.8 *(100,100, 0.8, 0.6, 0.2 , 0.2, -1*); *pro_NINL* =0.6 *(100,100, 0.4, 0.2, 0.6 , 0.2, -1*); *pro_NIL* =0.6 *(100,100, 0.4, 0.2, 0.2 , 0.6, -1*); $x_{LLM}$ =1 *(100,100, 0.4, 0.6, 0.2 , 0.2, 1*) ]. We finally plotted the trend curves of the four outcomes of each network dynamic scenario with the number of iterations, since the curves of *T*=300 and benchmark are basically the same among the four results, we only selected the results of the 100 iterations to show for a clearer representation, and the trend plots include (**A**) mean opinion value, which indicates the average opinion value of all agents for each iteration; (**B**) mean opinion change value , which denotes the absolute value of the difference between the opinion value of the agent at the current iteration and the opinion value at the previous iteration, and finally takes the average of all agents; (**C**) mean standard deviation, which denotes the standard deviation of all agents in each iteration, see Eq(7) in the method section; and (**D**) mean number of clusters, which denotes the number of hierarchical clusters of all agents in each iteration, see Eq(10) in the method section. In order to eliminate randomness of the four results, our values at iteration t are the average calculations of 100 repeated simulations of that network dynamics scenario at iteration t.

**Impacts on the evolution and formation of opinions.** Fig.3A presents our findings from the $NODE_{diff}$, $NODE_{conv}$, $NODE_{SD}$, and $NODE_{clus}$ sections. For $NODE_{diff}$, we observe that changes in agent opinions exhibit a significant positive correlation solely with the output value of LLM. No significant relationship exists with either the proportion of the three agents or the threshold value of the agents. In addition, the curve representing the change in opinion value influenced by the positive and negative LLM values is symmetric with respect to the y=0 axis of symmetry during iteration (Fig.3B). These findings suggest that LLM can facilitate educational efforts aimed at guiding the positive development of collective opinion. However, the influence relationship described above indicates that negative or toxic public opinions generated by LLMs can also lead to negative development of collective opinion networks. Therefore, intervention and control approaches need to be explored further to ensure the proper evolution of opinion networks. Additionally, Fig.3C shows that the standard deviation of the agents roughly obeys a parabolic distribution with $x_{LLM}$=0 as the symmetry axis and the smallest value here. This result suggests that as the LLM output value approaches the initial mean value of the opinion network, the final distribution of opinions becomes less heterogeneous.

In the simulations for $NODE_{conv}$, the opinion convergence times of NIN and NINL are both significantly and positively correlated with the threshold value, as the threshold value increases, the intensity of opinion interaction also increases, and the convergence time prolongs. However, the opinion convergence time of NIN is not significantly correlated with the proportion of the three agents or the output value of LLM, suggesting that the time required for the opinion of NIN in the population to reach a steady state is only related to the threshold. The convergence time of NINL exhibits a significant positive correlation with the NIN to NIL ratio, and a significant negative correlation with the NINL to NIN ratio. Conversely, the convergence time of all agents displays a significant positive correlation with the NIN to NINL ratio, and a significant negative correlation with the NIL to NIN ratio. These findings indicate that an increased number of individuals who do not employ LLMs within the opinion network result in extended convergence times for opinions held by those who partially rely on both LLMs and social groups to reach a stable state. A greater number of individuals who partially rely on LLMs lead to shorter convergence times for their individual opinions, but longer convergence times for collective opinions. Conversely, a great number of people who fully rely on LLM will increase the intensity of opinion interaction resulting in a long convergence time for NINL, but a short convergence time for collective opinions.

For $NODE_{SD}$, the standard deviations of opinions of NIN and NINL are both significantly and negatively correlated with the threshold value, indicating that a larger threshold value results in less discrete opinions compared to the mean value. The correlations between the standard deviation of NIN and NINL and the proportions of the three agents are more consistent. They are significantly and inversely correlated with the proportions of NIN and NINL, but significantly and positively correlated with the proportion of NIL. These findings suggest that a higher proportion of NIL is associated with increased disagreement within NIN and NINL, while a higher proportion of NIN and NINL is linked to greater convergence of their internal opinions. In contrast, the standard deviation of collective opinion is significantly positively correlated with the proportion of NINL, significantly negatively correlated with the proportion of NIL, and only slightly negatively correlated with the proportion of NIN. Additionally, Fig.3D shows that the standard deviation of the agents decreases very slowly until the threshold is less than 0.5 and decreases very quickly after the threshold is greater than 0.5. This result indicates that even with human intervention, the dispersion of the opinions of each agent is large as long as the threshold of the agent is below 0.5. Once the human intervention exceeds this threshold, the marginal effect would be significant, and the tendency to reach consensus would increase significantly.

For $NODE_{clus}$, the number of opinion clusters for NIN and NINL is negatively associated with the threshold value, showing that the larger the threshold value, the fewer the number of opinion clusters and the more opinions tend to be consensus-oriented. The number of clusters for NIN, NINL, and the group is significantly negatively correlated with the proportion of NIL. This observation suggests that a higher proportion of NIL not only results in a reduced number of clusters of both internal and collective opinions but also leads to a more consensual opinion among individuals who do not use LLMs or partially rely on LLMs. Additionally, the number of clusters of NIN is significantly positively correlated with the proportion of NIN and significantly negatively correlated with the proportion of NINL. This result indicates that the greater the proportion of NIN, the more dispersed the opinions within NIN. However, the number of clusters of NINL is significantly negatively correlated with the proportion of NIN and significantly positively correlated with the proportion of NINL, indicating that increasing the proportion of NINL will concentrate the opinions within NINL. Finally, Fig.3E shows that the number of opinion clusters decreases rapidly when the threshold is increased from 0 to 0.4 and then decreases slowly when the threshold is greater than 0.4. This result suggests that the initial threshold value must be set appropriately to achieve a balance between opinion diversity and consensus. Combining the results from Fig.3D and Fig.3E, we observe that increasing the threshold through intervention can quickly converge chaotic opinions into multiple distant opinion clusters when the threshold is less than 0.3. However, when the threshold is greater than 0.7, The number of opinion clusters is small, preferring to reach a consensus.

In summary, Fig.3A-E suggest that the overall convergence of the opinion network is slower and the opinion distribution is more divided when more people partially rely on LLM. In contrast, as the number of individuals who solely rely on LLMs increases, the convergence of the opinion network accelerates, and the opinion distribution becomes more concentrated and oriented towards consensus. Therefore, maintaining opinion diversity and full interaction of opinions requires a large proportion of NINL. However, excessive reliance on LLM can lead to a rapid convergence of opinion networks, which may limit opinion diversity and compromise the quality of collective opinion formation.

Fig.3F provides additional confirmation of the observations presented in Fig.3A and Fig.3B. Specifically, the minimum and maximum values of the opinion difference corresponded to parameter values of (0.95, 0.20, 0.80, 0.00, -1.00) and (0.93, 0.19, 0.81, 0.00, 1.00), respectively. Notably, the output values of LLMs for these two sets of parameters were diametrically opposed. Fig.3G indicates that to achieve a rapid attainment of a stable state in collective opinion exchange, individual cognitive acceptability should be close to 0, the output opinion value of LLMs should be 0, and the proportions of NIN, NINL, and NIL agents should be approximately 27%, 27%, and 46%, respectively. For a more intense opinion exchange process, the individual cognitive acceptability should preferably be 0.6, the output opinion value of LLMs should be close to 0, and the proportions of NIN, NINL, and NIL agents should be approximately 44%, 41%, and 15%, respectively. Fig.3H illustrates that the minimum value of the standard deviation of collective opinion occurs when the fraction of NIL is equal to 1 (0.50, 0.00, 0.00, 1.00, 0.00), and individual opinions are consistent with the output of LLMs and do not change. In contrast, the standard deviation of collective opinion reaches its maximum value when the acceptability of individuals is 0.13, and the output of LLMs is 0.20, with the proportion of three nodes roughly 37%, 31%, and 32%. Fig.3I demonstrates that when collective opinion reaches a consensus, the acceptability of individuals is 0.14, the output opinion value of LLMs is 0, and the proportions of the three agents are roughly 27%, 27%, and 46%. Conversely, when collective opinion reaches polarization, the acceptability of individuals is 0.92, the output opinion value of LLMs is 0, and the proportions of the three agents are roughly 35%, 34%, and 31%. Finally, when collective opinion reaches maximum fragmentation, the acceptance of individuals is 0, the output opinion value of LLMs is 0.06, and the proportions of the three agents are approximately 23%, 71%, and 6%.

In general, Fig.3F-I provide a different perspective from Fig.3A-E in that they equilibrate five parameters and provide optimal solutions for different needs in opinion networks involving LLMs, e.g., one wants to maximize opinion diversity, then it is better to have 70% of the population using LLMs and 20% of the population not using LLMs. In addition, our results in this section can provide the direction of intervention for other parameters when some parameter is known as a priori knowledge.

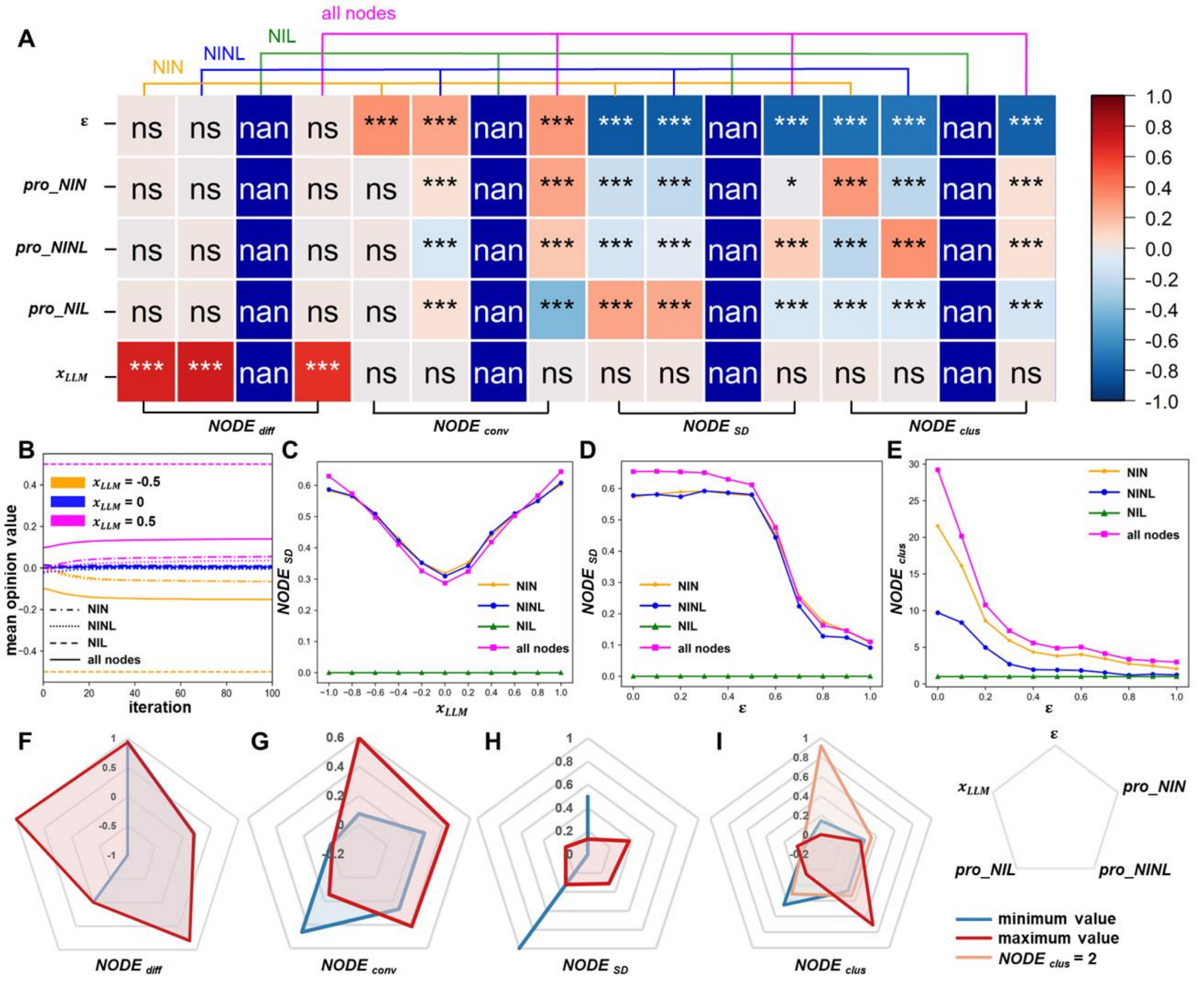

**Fig.3.** Relationships of five filtered parameters and four indicators. We aim to investigate the precise trend of the impact of different parameter settings on the opinion network, so we divide the values of each parameter into 11 cases, i.e., the possible values of $\varepsilon$, *pro_NIN*, *pro_NINL*, and *pro_NIL* are [0,0.1,0.2,0.3,0.4,0.5,0.6,0.7,0.8, 0.9,1] and the sum of the proportions of the three agents is 1, the possible values of $x_{LLM}$ take the values [-1,-0.8,-0.6,-0.4,-0.2,0,0.2,0.4,0.6,0.8,1], and there are a total of 7986 combinations of parameters for the opinion dynamics after traversing all cases. We then calculate the values of the four indicators ($NODE_{diff}$, $NODE_{conv}$, $NODE_{SD}$, and $NODE_{clus}$) for each combination of agents with three different use strategies and for all agents in the network, the detailed description and computation process of each indicator is described in the methods section. To eliminate randomness of our results, we repeat the simulation 100 times for each combination, and the final indicator values are the average of the results of these 100 times. (**A**) Pearson correlation coefficients of 5 filtered parameters and 4 indicators for the agents with different usage strategies. The Pearson correlation coefficients takes values in the range [-1, 1], with a value of 0 indicating that the two variables are not correlated, a positive value indicating a positive correlation, and a negative value indicating a negative correlation, the different colors of the squares in the figure represent different values of the Pearson correlation coefficients, and the legend is on the right, Since the value of the NIL does not change, resulting in the values of its four indicators not changing either, none of their Pearson correlation coefficients with the five parameter values exist. We also conducted t-test to test the degree of significance of the Pearson correlation coefficients, * denotes the degree of significance, i.e., the P-value, ***$P$<0.001, **$P$<0.01, *$P < 0.05$, ns means $P \geq 0.05$, nan means none $P$ value exist. In order to derive detailed trends that were not available from the correlation analysis, we then use the benchmark scenario in Fig.2, and keep changing the value of only one of the parameters at a time, and repeat the simulation 100 times to plot detailed trend of the indicators with respect to the parameters, some additional findings we obtained are shown in (**B**), (**C**), (**D**) and (**E**). To obtain the optimal combination of parameters for different use and intervention strategies of LLMs, we computed the average values

of the five parameters when the minimum or maximum values are obtained for the four indicators, respectively. Which are shown in (**F**), (**G**), (**H**) and (**I**). In order to make the results more robust, when different combinations of parameters reach extreme values at the same time, we select all combinations of parameters that reach extreme values and calculate their average values, and when this does not happen, we select the top ten combinations of parameters to calculate the average value. The minimum value of the opinion clusters in (I) is 1, which represents the consensus of opinions, and we additionally show the case where the mean value of the opinion clusters is 2, which represents the polarization of opinions in reality.

**Countered solutions and intervention strategies to hazards of LLMs.** Fig.4A shows that randomly adding agents with opposite, neutral, or random opinions significantly improves scenarios with negative LLMs' values, such as bias. There is no significant difference between these three approaches. However, the final opinion distribution for the approach introducing opposite values spans a larger range than the remaining two approaches, with the minimum value still larger than the original scenario, and the maximum value a large improvement over the original scenario. This measure is more suitable for scenarios that require a complete and rapid reversal of opinion, such as correcting bias as opposed to a certain established fact. Approaches that introduce neutral and random agents have the smallest final opinion distribution span and significantly improve the minimum value of opinion, but not the maximum value. These two approaches are more robust and more suitable for scenarios that require slow changes in population misperceptions, such as biases against race, poverty, and disability.

In Fig.4B, we observe that the standard deviation of individuals in the group who are all partially dependent (0.370) on LLMs is significantly larger in the extreme scenario than in the non-dependent (0.267) and fully dependent scenarios (0). In Fig.4C, we observe that the number of opinion clusters in the group that are all non-dependent or partially dependent on LLMs is significantly larger in the extreme scenario than in the fully dependent scenario. These findings further confirm that LLMs can be effective in increasing opinion diversity if used appropriately, whereas complete reliance on LLM leads to a stronger collective consensus.

Fig.4D illustrates different intervention strategies for using LLMs in different contexts with specific objectives of collective opinion difference, convergence, and fragmentation. Overall, we find that when our desired collective opinion and LLMs' output values are opposite, the most effective method for human intervention is to randomly introduce three types of agents. When our desired collective opinion and LLMs' output values are not opposite, the most effective method to change the collective opinions is to vary the proportion of people who use different strategies. Increasing the proportion of NINL, NIN and the threshold of the population can effectively intensify the interactions of collective opinions. Increasing the proportion of NINL can effectively diversify collective opinion. There are many specific implementation options for each intervention method in Fig.4D, and we provide some common options based on proven psychological and sociological knowledge in the Conclusion section.

In summary, the experiments and results in this section provide valuable insights into the effective use of LLMs in shaping collective opinion formation and convergence. The findings highlight the importance of the appropriate use of LLMs in promoting opinion diversity and fragmentation-oriented opinion formation. These results have important implications for the design and implementation of effective interventions aimed at promoting the positive development of opinion networks in various contexts.

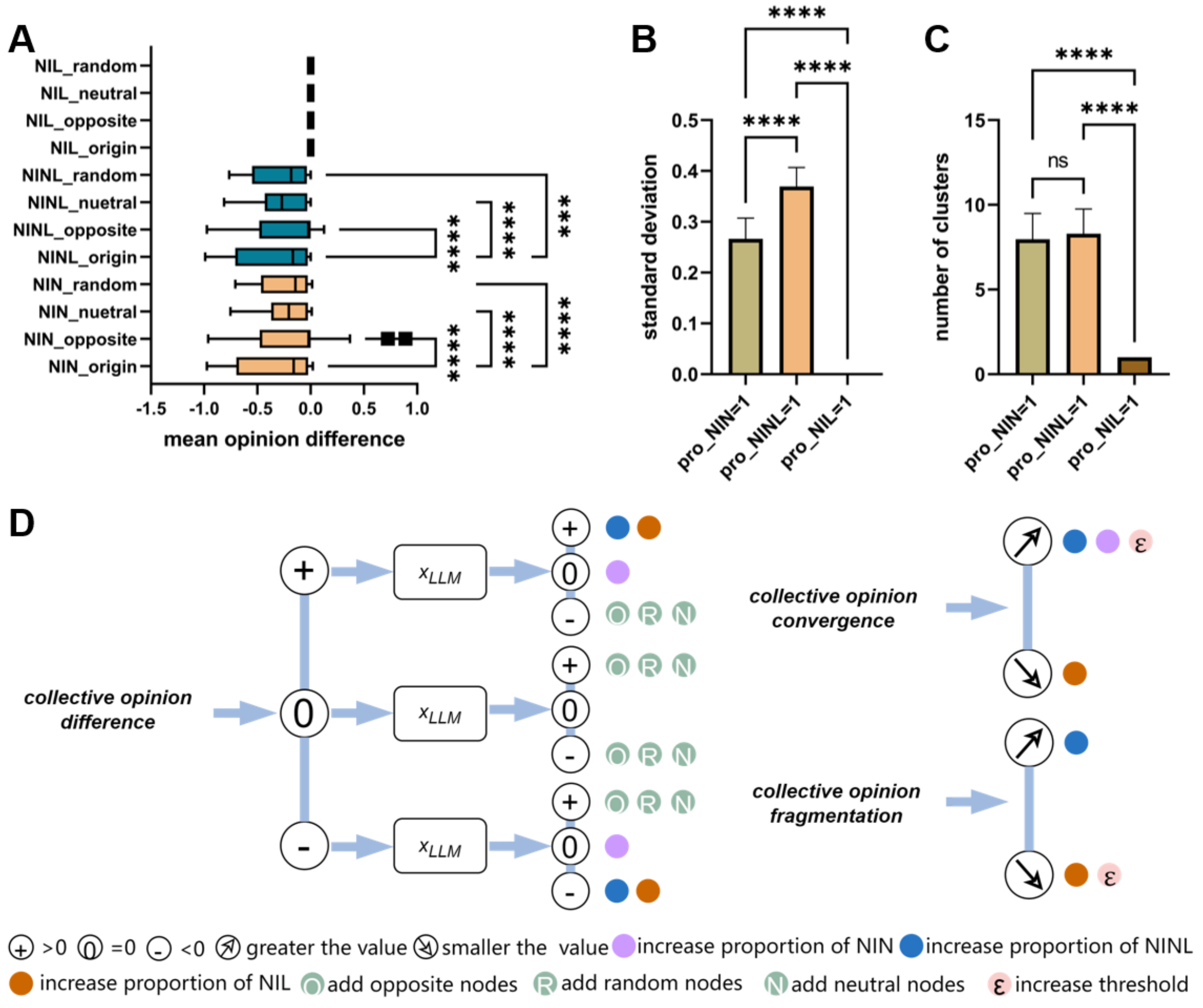

**Fig.4.** Countered solutions and intervention strategies to hazards of LLMs. (**A**) The distribution of the mean opinion difference of three categories of agents in original scenario and 3 countered scenarios. In the context of direct or implicit bias and toxicity of LLM, how to intervene reasonably and effectively in the opinion network is important to eliminate social discrimination and verbal violence, etc. According to Fig.3A, we found that only the $x_{LLM}$ has a significant relationship with the $NODE_{diff}$, therefore, in the existing framework, there is a lack of means to address the change in the opinion network caused by the output opinion values of LLM. To address this issue, three attempts were made. Specifically, we first select all $NODE_{diff}$ values (N=726) when the $x_{LLM}$ is -1 as the resultant values of the original dynamic network here, which were calculated in Fig.3A. We then introduce three countered solutions, which are 1) agents of opposite opinions (here i.e. 1); 2) agents of neutral opinions (here i.e. 0 ); and 3) agents of random opinions (with the value domain [-1,1]) added randomly in iterations, and once they were added, they all keep the initial opinion values, the probability of adding agents at iteration t is all 0.1, the number of agents added is all 2, and the LLM value is fixed at -1, too, all other possible parameter combinations (N=726) are traversed, and each combination of parameters is simulated 100 times, the average value is taken as the final result corresponding to each parameter combination. The distribution of the mean opinion difference were plotted using a box plot, with the five vertical lines from left to right indicating the minimum, lower quartile, median, upper quartile and maximum of the data, respectively, We also conducted one-way ANOVA analysis to investigate whether the difference between the two groups of data was significant, we only show here the p-value less than 0.05, * denotes the degree of significance, i.e., the P-value, ****$P$<0.0001, ***$P$<0.001, **$P$<0.01, *$P$<0. 05. (**B**) The distribution of the standard deviation with *pro_NIN*, *pro_NINL*, *pro_NIL* equal to 1, respectively. In three extreme usage strategies here, we select all $NODE_{SD}$ values (N=121) which were calculated in Fig.3A., the height of the bar chart represents the mean standard deviation, the error bars show the 95% confidence interval of the statistical data, the test method for difference significance is the same as A. (**C**) The distribution of the number of clusters with *pro_NIN*, *pro_NINL*, *pro_NIL* equal to 1, respectively. In three extreme usage strategies here, we select all $NODE_{clus}$ values (N=121) which were calculated in Fig.3A., the symbols in the figure are the same as in (B). (**D**) Appropriate intervention strategies of LLMs meeting different scenarios. We take the different needs in reality into account, and plot this diagram based on the correlation matrix and the results of the above three diagrams from three aspects: collective opinion difference, collective opinion convergence time and collective opinion fragmentation.

## Discussion

In the past decades, the study of social opinion evolution has attracted much attention[43,44]. A number of opinion models have been proposed based on well-studied sociological and psychological principles[45-49], such as the DeGroot model[50], voter model[51], Sznajd model[52], Friedkin and Johnsen model[53], and bounded confidence model[54]. A lot of effort has been put into understanding opinion dynamics on the traditional social interactions media era, but little research has been done on opinion models to incorporate the interaction patterns of LLMs. Unlike traditional media interaction, LLMs interact with users in both directions, and LLMs are more likely to output toxic and biased content[25-27], so the potential impact of LLMs on opinions is not fully understood. Researchers have identified six specific risk areas of LLMs[26], including the potential for implicit bias (e.g., gender, race, and country) and the risk of indulging in their use, which were highly relevant to opinion dynamics. Recent research also have confirmed that uncensored LLMs can significantly influence individual opinions[38,41]. On the basis of the above study, we conducted millions of simulations with a modified HK model to dig into the impact of LLMs on social opinion dynamics and propose different target-oriented interventions for the utilization of LLMs.

Our results show that a broader cognitive acceptability leads to an eventual consensus of collective opinion, which is consistent with previous findings on opinion dynamics[55-57]. Deaprting from this point, we demonstrate that the marginal impact of threshold on collective opinion formation is nonlinear. Specifically, thresholds less than 0.5 contribute rapidly to the decentralized aggregation of opinions, thresholds greater than 0.5 contribute rapidly to the overall consensus of opinions. This finding can enrich the theory of cognitive acceptability in opinion dynamics with the involvement of LLMs. The output opinion values of LLMs have a significant and positive effect on the formation of opinions. The use strategy of LLMs has a significant impact on the convergence and distribution of opinions. Moreover, an interesting phenomenon that the use strategies of partial and full reliance on LLMs lead to almost exactly opposite effects on the convergence and distribution of opinions is observed, which may be linked to the multiple sources of opinion exchange that partially rely on LLMs.

We propose coping strategies for the problems that have been demonstrated for LLMs at this stage: bias and toxicity. We find that introducing agents randomly in the network with opposite/neutral/random opinions all significantly reduce the tendency of overall opinion negativization, and that the latter two approaches are more robust. We also explore relevant interventions for the potential risk of overdependence based on the correlation matrix, such as converting individuals who are overly dependent on LLMs to partial dependence. Our study provides different target-oriented strategies for the use and intervention of LLMs, they mainly increase the threshold of individuals, increase the proportion of NIN/NINL/NIL, and add opposite /neutral/random agents. There are many different implementation options for different use and intervention strategies. For example, to increase the threshold of individuals, we can improve education and promote intercultural competence[58]; to increase the proportion of NINL, i.e., encourage the rationalized use of LLMs, we can improve people's technological literacy through various forms of publicity, education, and training activities, so that they can understand the advantages and disadvantages of big language models, and we also need to simultaneously develop relevant regulatory and normative measures to protect user data privacy, and avoid problems of model abuse and over-reliance.

To the best of our knowledge, our study investigates for the first time the impact of LLMs on opinion dynamics, starting from different output opinion values of LLMs and different usage strategies. Our findings help promote a deeper public understanding of the influence of LLMs on opinion evolution and support groups and individuals in choosing the best usage strategies with respect to their own demands. Our study highlights that rationalizing the use of LLMs can significantly increase the diversity of opinions compared to not using them, but over-reliance can lead to the opposite situation. Despite the current prohibition of LLMs in numerous companies, schools, and organizations, our findings provide compelling evidence for the rational use of LLMs. Our study supports the notion that such use should not impede the development of artificial general intelligence (AGI) technologies, including LLMs, while also delivering additional benefits to both personal and professional spheres, such as improved productivity and diversified communication channels. Our study offers novel insights into the intricacies of public opinion dynamics in the era of LLMs. Moreover, our findings support the urgent request of theoretical studies of the evolution and formation of opinions in the age of LLMs and even in the future age of AGI.

Our analysis provides new insights for general, archetypal opinion dynamics simulating approaches. It has however some limitations since realistic opinion interaction situations are far more complex than theoretical models. Our study considers a number of phenomena in realistic opinion dynamics, including the influence of authority, the influence of stubbornness, and the influence of arbitrary events outside the network. However, the mechanisms by which people express and exchange their opinions in reality are more complex. For example, some people express their opinions differently from their internal opinions, some people will amplify their opinions when expressing them, and even some people express different opinions to different people at the same time, on the same topic.

## Methods

**Opinion dynamics model.** We first take into account that the opinion values in the opinion network are continuous rather than discrete variables like voting questions[59], and therefore choose a continuous opinion model rather than a discrete opinion model like the voter model[51]. Moreover, an agent will neither simply share nor completely disregard the opinions of other agents but will take these opinions into account to a certain extent in forming his/her new opinions in a process defined by a fusion rule. Hence, rather than DeGroot model[50], we choose to base our model on the classical Hegselmann-Krause (HK) model[42], which is one of the most widely used opinion models of the bounded confidence model[54,60], moreover, after taking into account influence of LLMs and complex realities, we propose the new opinion model. The classical HK model[42] is defined as:

$$x_j(t+1) = \left|J\big(j, x(t)\big)\right|^{-1} \sum_{i \in J(j,x(t))} x_i(t)\,, for\ t \in T \tag{1}$$

Where $J(j,x) = \{1 \le i \le n \,|\, |x_i(t) - x_j(t)| < \varepsilon_j\}$ and $\varepsilon_j$ is the confidence level of agent $j$. $x_i(t)$ is the opinion value of Agent $i$ at time *t*. Agent $j$ takes only those agents $i$ into account whose opinions differ from his own by not more than $\varepsilon_j$. The base case assumes a uniform level of confidence, i.e., $\varepsilon_j = \varepsilon$ for all agents $j$.

Compared to the classical HK model, we first take the different usage strategies of LLMs into account, categorize agents into three categories, NIN, NINL and NIL, which also indicates the agents are influenced in different extent: NIN represents agent who does not use LLMs, and is completely unaffected directly by LLMs and only influenced by neighboring nodes; NINL represents agent who partially rely on LLMs, and is influenced directly by both LLMs and neighboring nodes; NIL represents agent who totally rely on LLMs, and is completely influenced by LLMs and is not influenced by neighboring nodes. We then take the complex realities into account, propose three modifications: a) The authoritative effect is simulated by taking into account the different levels of authority and influence of different nodes, instead of each neighbor within the agent threshold having the same weight; b ) The stubbornness of different agent is simulated by randomly introducing the stubbornness degree into the updating formula of agent opinions; c) The influences of arbitrary events to opinions are simulated by introducing a random event in each opinion iteration, and it randomly affects some of these agents.

We adopted Erdos-Renyi graph to conduct our experiments, as it is most commonly used in existing literature to model social networks[61]. Given a random full-connected opinion networks (G) in which the number of group size is *N*, and the three categorized nodes occupy different proportions. Let $x_i(t)$ represent the opinion of agent $i$ at time $t$ , and its value range is [-1, 1], a value of $-1.0$ means a very negative opinion, and a value of 1 means a very positive opinion.; let $au_i$ represent the authority of agent $i$, which equals to the number of its neighbors divided by the number of nodes other than itself; the given confidence level for agent $i$ is $\varepsilon_i$, the given stubborn degree for agent $i$ is $sd_i$.The value ranges of $au_i$, $\varepsilon_i$ and $sd_i$ is [0, 1]. The initial value of $x_i(t)$ and $sd_i$ are randomly assigned and obey uniform distributions respectively in their value ranges. The number of opinion exchanges, i.e., the total iteration number for opinion dynamics is *T*. The updating formulas of these three categories of nodes are as follows.

a) For NIN:

$$\begin{cases} \quad if \ \exists x_i(t), |x_i(t) - x_j(t)| \le \varepsilon_j \\ then \ \ x_j(t+1) = x_j(t) * sd_j + \left\{ \sum_{i \in J(j,x(t))} \frac{au_i}{\sum_{i \in J(j,x(t))} au_i} x_i(t) \right\} * (1 - sd_j) \end{cases} \tag{2}$$

Where $J(j,x) = \{1 \le i \le n \,|\, |x_i(t) - x_j(t)| \le \varepsilon_j\}$.

b) For NINL:

$$\begin{cases} if \quad |x_j(t) - x_{LLM}(t)| > \varepsilon_j \cap \ \exists x_i(t), |x_i(t) - x_j(t)| \le \varepsilon_j \\ then \ \ x_j(t+1) = x_j(t) * sd_j + \left\{ \sum_{i \in J(j,x(t))} \frac{au_i}{\sum_{i \in J(j,x(t))} au_i} x_i(t) \right\} * (1 - sd_j) \\ if \quad |x_j(t) - x_{LLM}(t)| \le \varepsilon_j \cap \ \exists x_i(t), |x_i(t) - x_j(t)| \le \varepsilon_j \\ then \ \ x_j(t+1) = x_j(t) * sd_j + \left\{ \begin{array}{l} \sum_{i \in J(j,x(t))} \frac{au_i}{\sum_{i \in J(j,x(t))} au_i + au_{LLM}} x_i(t) \\ + \frac{au_{LLM}}{\sum_{i \in J(j,x(t))} au_i + au_{LLM}} \cdot x_{LLM}(t) \end{array} \right\} * (1 - sd_j) \end{cases} \tag{3}$$

Where $J(j,x) = \{1 \le i \le n \big| |x_i(t) - x_j(t)| \le \varepsilon_j\}$ , $x_{LLM}(t)$ is the opinion value delivered by LLM at time $t$ , as it is obtained by automatic text generating based on a large amount of historical data, individuals have a negligible impact on the output of LLMs when they interact with them in a Q&A format. We thus assume it as a constant during each iteration in this study, i.e., $x_{LLM}(t)$= $x_{LLM}$ for all times $t$. $au_{LLM}$ is the authority of LLM, we treat it as 1 by the assumption that LLM has the potential to connect every agent.

c) For NIL:

$$x_j(t+1) = x_{LLM}(t) \tag{4}$$

In general, our model's single simulations are performed according to the above rules, and to reduce the randomness of the results, we repeat the simulations one hundred times with the same controlled parameters. Our model has seven controlled parameters, and they are the number of group size (*N*), the number of opinion exchange (*T*), the cognitive acceptability of agents ($\varepsilon$), the proportion of NIN (*pro_NIN*), the proportion of NINL (*pro_NINL*), the proportion of NIL (*pro_NIL*), opinion value of LLM ($x_{LLM}$).

**Multi-simulation post-processing method.** We aim to investigate the precise trend of the impact of different parameter settings on the opinion network, which means that the steps of our parameters will be more intensive, so the combination of the above seven parameters may appear in hundreds of millions of different scenarios. Considering the computational efficiency and time of the model, we first filtered out the five parameters that have the greatest impact on the results by the set baseline scenarios. We then delve into the impact of LLMs on opinion networks in terms of opinion evolution and opinion distribution by performing 100 simulations under every combination of selected parameters. Opinion evolution refers to the evolution pattern of node opinion values, including two indicators: opinion difference and opinion convergence time; opinion distribution refers to the distribution of node opinion values, including two indicators: opinion standard deviation difference and the number of opinion clusters. The detailed description and calculation methods of the above four indicators are as follows.

Opinion difference is the evolution of the value of an agent's opinion on a topic. In this study, we categorized three types of nodes and computed their mean opinion difference, with negative values indicating that the mean opinion of that type of node undergoes a negative change, i.e., becomes more negative, and positive values the opposite. The formula of mean opinion difference ($NODE_{diff}$) is as follows.

$$NODE_{diff} = \frac{\sum_{s\in S}\sum_{i=1}^{n}\left(x_i(T) - x_i(0)\right)}{Sn} \tag{5}$$

Where $S$ represents the number of simulations, $n$ represents the number of specific category of nodes, $x_i(T)$ and $x_i(0)$ represent the final value and initial value of node $i$.

Opinion convergence time is the timestep it takes for an agent's opinion to evolve to a stable state. In this study, we categorize three types of nodes and compute their average opinion convergence time. The larger the value, the longer the average opinion convergence time of that type of node, i.e., the longer it takes for the opinion o to reach a stable state, and the more intense and chaotic the interaction process of opinions. The formula of mean opinion convergence time ($NODE_{conv}$) is as follows.

$$NODE_{conv} = \frac{\sum_{s\in S}(t|\ \forall i, |x_i(t) - x_i(t-1)| \le \tau)}{S} \tag{6}$$

Where $(t|\ \forall i, |x_i(t) - x_i(t-1)| \le \tau)$ means for all nodes belonging to the same specific category of nodes, if the difference between their value at time $t$ and their value at time $t-1$ is less than $\tau$, We take $\tau$ to be five thousandths of 1, i.e. 0.005, the time $t$ is taken as their convergence time.

Opinion standard deviation is the degree of dispersion of a group's opinions relative to the mean value. In this study, we categorize three types of nodes and compute their average opinion standard deviation. The larger the value, the more discrete the overall distribution of opinions of the nodes is relative to the mean, i.e., the wider the overall distribution of opinions. The formula of mean opinion standard deviation ($NODE_{SD}$) is as follows.

$$NODE_{SD} = \frac{\sum_{s\in S}\sqrt{\frac{\sum_{i=1}^{n}\left(x_i(T) - \bar{x}(T)\right)^2}{n-1}}}{S} \tag{7}$$

Where $\bar{x}(T)$ represents the mean final value of a specific category of nodes.

The number of opinion clusters can effectively compensate for the lack of standard deviation in portraying data distribution, for example, the standard deviation is large when the opinions are polarized, but the opinions are concentrated. Therefore, we introduce the number of opinion clusters to indicate the degree of opinion aggregation. The larger the value, the more points the opinion distribution of the node is concentrated, i.e., the more the overall opinion distribution tends to be split, and a value of 2 indicates that the opinion distribution of the node is polarized, a value of 1 indicates that the opinion distribution of the node is consensus.

The commonly used K-means clustering method needs to specify the number of categories k in advance, which obviously does not meet our needs because the final distribution of opinions cannot be specified in advance, and density-based clustering methods, such as DBSCAN, do not take well into account the fragmentation of opinion, so we apply single linkage hierarchical clustering, which can compensate the defects of the above two clustering methods, and is an agglomerative clustering algorithm that builds trees in a bottom-up approach[62,63]. Specifically, we first take the $x_i(T)$ obtained from a single simulation, i.e., the final opinion value of each agent, as a separate cluster $C_i$, and then calculate the distance between clusters using the Manhattan distance (see Eq.(8)), followed by merging the two clusters with the closest distance into a new cluster, and the distance between the new merged cluster and the other cluster is the distance between the sample points with the smallest distance in the two clusters (see Eq.(9)), and keep repeating the merged clusters until all agents' final opinion values are assigned to the one single cluster. After obtaining the final dendrogram, we conduct several trials to choose to cut the dendrogram at height 0.2, which implies that the radius of the resulting clusters cannot exceed 0.2, and we traverse from top to bottom and count the number of clusters that satisfy the condition as number of opinion clusters, after 100 simulations, we obtain mean number of opinion clusters $(NODE_{clus})$ (see Eq. (10)).

$$dist_{ij} = dist(C_i, C_j) = |x_i(T) - x_j(T)| \tag{8}$$

Where $dist_{ij}$ represents the Manhattan distance between initial clusters $C_i$ and $C_j$.

$$dist_{CD} = min(dist(C_A, C_D), dist(C_B, C_D)) = min(|x_c - x_d|), c \in C_C, d \in C_D \tag{9}$$

Where $dist_{CD}$ represents the Manhattan distance between clusters $C_C$ and $C_D$, clusters $C_A$ and $C_B$ are the results of last clustering, $C_C = C_A + C_B$ is the results of present clustering, which means the agent in $C_C$ is the concatenation of the agents in $C_A$ and $C_B$.

$$NODE_{clus} = \frac{\sum_{s \in S}(m| \forall p, dist(p, p+1) \le \delta)}{S} \tag{10}$$

Where $(m| \forall i, dist(i, i+1) \le \delta)$ represents traversing the dendrogram from top to bottom, and return the number of clusters $m$ when the distance between all adjacent clusters (i.e., $p$ and $p+1$) is less than $\delta$ for the first time, we take $\delta$ as one tenth of the value range of $x_i(t)$, i.e., 0.2.

**Acknowledgements**

This study was supported by the National Natural Science Foundation of China (#71971196).

**Reference**

1 Centola, D., Becker, J., Brackbill, D. & Baronchelli, A. Experimental evidence for tipping points in social convention. *Science* **360**, 1116-1119, doi:10.1126/science.aas8827 (2018).

2 Aririguzoh, S. Communication competencies, culture and SDGs: effective processes to cross-cultural communication. *Humanities and Social Sciences Communications* **9**, 96, doi:10.1057/s41599-022-01109-4 (2022).

3 Li, F., Liu, Y. & Meng, T. Discursive strategy of opinion expression and government response in China: Text analysis based on online petitions. *Telematics and Informatics* **42**, 101238, doi:https://doi.org/10.1016/j.tele.2019.06.001 (2019).

4 Muchnik, L., Aral, S. & Taylor, S. J. Social Influence Bias: A Randomized Experiment. *Science* **341**, 647-651, doi:10.1126/science.1240466 (2013).

5 Perra, N. & Rocha, L. E. C. Modelling opinion dynamics in the age of algorithmic personalisation. *Scientific Reports* **9**, 7261, doi:10.1038/s41598-019-43830-2 (2019).

6 Paluck, E. L., Shepherd, H. & Aronow, P. M. Changing climates of conflict: A social network experiment in 56 schools. *Proceedings of the National Academy of Sciences* **113**, 566-571, doi:10.1073/pnas.1514483113 (2016).

7 Ferraz de Arruda, G., Petri, G., Rodriguez, P. M. & Moreno, Y. Multistability, intermittency, and hybrid transitions in social contagion models on hypergraphs. *Nature Communications* **14**, 1375, doi:10.1038/s41467-023-37118-3 (2023).

8 Proskurnikov, A. V. & Tempo, R. A tutorial on modeling and analysis of dynamic social networks. Part I. *Annual Reviews in Control* **43**, 65-79, doi:https://doi.org/10.1016/j.arcontrol.2017.03.002 (2017).

9 Proskurnikov, A. V. & Tempo, R. A tutorial on modeling and analysis of dynamic social networks. Part II. *Annual Reviews in Control* **45**, 166-190, doi:https://doi.org/10.1016/j.arcontrol.2018.03.005 (2018).

10 Hassani, H. *et al.* Classical dynamic consensus and opinion dynamics models: A survey of recent trends and methodologies. *Information Fusion* **88**, 22-40, doi:https://doi.org/10.1016/j.inffus.2022.07.003 (2022).

11 Li, L., Fan, Y., Zeng, A. & Di, Z. Binary opinion dynamics on signed networks based on Ising model. *Physica A: Statistical Mechanics and its Applications* **525**, 433-442, doi:https://doi.org/10.1016/j.physa.2019.03.011 (2019).

12 Laptev, A. A. Modeling of Social Processes Based on T.Parsons Ideas. *Advances in Complex Systems* **03**, 99-106, doi:10.1142/S021952590000008X (2000).

13 Weisbuch, G., Deffuant, G., Amblard, F. & Nadal, J.-P. Meet, discuss, and segregate! *Complexity* **7**, 55-63, doi:https://doi.org/10.1002/cplx.10031 (2002).

14 Borkar, V. S. & Reiffers-Masson, A. Opinion Shaping in Social Networks Using Reinforcement Learning. *IEEE Transactions on Control of Network Systems* **9**, 1305-1316, doi:10.1109/TCNS.2021.3117231 (2022).

15 Noorazar, H. Recent advances in opinion propagation dynamics: a 2020 survey. *The European Physical Journal Plus* **135**, 521, doi:10.1140/epjp/s13360-020-00541-2 (2020).

16 Xiong, F., Liu, Y., Wang, L. & Wang, X. Analysis and application of opinion model with multiple topic interactions. *Chaos: An Interdisciplinary Journal of Nonlinear Science* **27**, 083113, doi:10.1063/1.4998736 (2017).

17 Zhang, N., Huang, H., Su, B., Zhao, J. & Zhang, B. Information dissemination analysis of different media towards the application for disaster pre-warning. *PloS one* **9**, e98649, doi:10.1371/journal.pone.0098649 (2014).

18 Kubin, E. & von Sikorski, C. The role of (social) media in political polarization: a systematic review. *Annals of the International Communication Association* **45**, 188-206, doi:10.1080/23808985.2021.1976070 (2021).

19 Flamino, J. *et al.* Political polarization of news media and influencers on Twitter in the 2016 and 2020 US presidential elections. *Nature Human Behaviour* **7**, 904-916, doi:10.1038/s41562-023-01550-8 (2023).

20 Fan, C., Jiang, Y., Yang, Y., Zhang, C. & Mostafavi, A. Crowd or hubs: Information diffusion patterns in online social networks in disasters. *International Journal of Disaster Risk Reduction* **46**, 101498, doi:10.1016/j.ijdrr.2020.101498 (2020).

21 Fan, C., Jiang, Y. & Mostafavi, A. Emergent social cohesion for coping with community disruptions in disasters. *Journal of the Royal Society Interface* **17** (2020).

22 Strömbäck, J. *et al.* News media trust and its impact on media use: toward a framework for future research. *Annals of the International Communication Association* **44**, 139-156, doi:10.1080/23808985.2020.1755338 (2020).

23 Yang, Y. *et al.* Exploring the emergence of influential users on social media during natural disasters. *International Journal of Disaster Risk Reduction* **38**, 101204, doi:https://doi.org/10.1016/j.ijdrr.2019.101204 (2019).

24 Aiyappa, R., An, J., Kwak, H. & Ahn, Y.-Y. Can we trust the evaluation on ChatGPT? *ArXiv* **abs/2303.12767** (2023).

25 Zhao, W. X. *et al.* A Survey of Large Language Models. *ArXiv* **abs/2303.18223** (2023).

26 Weidinger, L. *et al.* Ethical and social risks of harm from Language Models. *ArXiv* **abs/2112.04359** (2021).

27 Luitse, D. & Denkena, W. The great Transformer: Examining the role of large language models in the political economy of AI. *Big Data & Society* **8**, 20539517211047734, doi:10.1177/20539517211047734 (2021).

28 Pegoraro, A., Kumari, K., Fereidooni, H. & Sadeghi, A.-R. To ChatGPT, or not to ChatGPT: That is the question! *ArXiv* **abs/2304.01487** (2023).

29 Eysenbach, G. The Role of ChatGPT, Generative Language Models, and Artificial Intelligence in Medical Education: A Conversation With ChatGPT and a Call for Papers. *JMIR Med Educ* **9**, e46885, doi:10.2196/46885 (2023).

30 Bostrom, N. Information Hazards: A Typology of Potential Harms from Knowledge. **10** (2012).

31 Craft, J. T., Wright, K. E., Weissler, R. E. & Queen, R. M. Language and Discrimination: Generating Meaning, Perceiving Identities, and Discriminating Outcomes. *Annual Review of Linguistics* **6**, 389-407, doi:10.1146/annurev-linguistics-011718-011659 (2020).

32 McKee, K., Bai, X. & Fiske, S. *Understanding Human Impressions of Artificial Intelligence*. (2021).

33 Talboy, A. N. & Fuller, E. Challenging the appearance of machine intelligence: Cognitive bias in LLMs. *ArXiv* **abs/2304.01358** (2023).

34 West, J. D. & Bergstrom, C. T. Misinformation in and about science. *Proceedings of the National Academy of Sciences* **118**, e1912444117, doi:10.1073/pnas.1912444117 (2021).

35 Skitka, L. J., Mosier, K. L. & Burdick, M. Does automation bias decision-making? *International Journal of Human-Computer Studies* **51**, 991-1006, doi:https://doi.org/10.1006/ijhc.1999.0252 (1999).

36 Piloto, L. S., Weinstein, A., Battaglia, P. & Botvinick, M. Intuitive physics learning in a deep-learning model inspired by developmental psychology. *Nature Human Behaviour* **6**, 1257-1267, doi:10.1038/s41562-022-01394-8 (2022).

37 Smith, B. C. *The Promise of Artificial Intelligence: Reckoning and Judgment*. (The MIT Press, 2019).

38 Salewski, L., Alaniz, S., Rio-Torto, I., Schulz, E. & Akata, Z. In-Context Impersonation Reveals Large Language Models' Strengths and Biases. *ArXiv* **abs/2305.14930** (2023).

39 Ousidhoum, N. D., Zhao, X., Fang, T., Song, Y. & Yeung, D.-Y. in *Annual Meeting of the Association for Computational Linguistics.*

40 Venkit, P., Gautam, S., Panchanadikar, R., Huang, T.-H. K. & Wilson, S. in *Conference of the European Chapter of the Association for Computational Linguistics.*

41 Rutinowski, J., Franke, S., Endendyk, J., Dormuth, I. & Pauly, M. The Self-Perception and Political Biases of ChatGPT. *ArXiv* **abs/2304.07333** (2023).

42 Hegselmann, R. & Krause, U. Opinion dynamics and bounded confidence: models, analysis and simulation. *J. Artif. Soc. Soc. Simul.* **5** (2002).

43 Peralta, A. F., Kertész, J. & Iñiguez, G. Opinion dynamics in social networks: From models to data. *arXiv preprint arXiv:2201.01322* (2022).

44 Anderson, B. D., Dabbene, F., Proskurnikov, A. V., Ravazzi, C. & Ye, M. Dynamical networks of social influence: Modern trends and perspectives. *IFAC-PapersOnLine* **53**, 17616-17627 (2020).

45 Eisenberger, N. I., Lieberman, M. D. & Williams, K. D. Does rejection hurt? An fMRI study of social exclusion. *Science* **302**, 290-292 (2003).

46 Zhao, Y., Kou, G., Peng, Y. & Chen, Y. Understanding influence power of opinion leaders in e-commerce networks: An opinion dynamics theory perspective. *Information Sciences* **426**, 131-147 (2018).

47 Dandekar, P., Goel, A. & Lee, D. T. Biased assimilation, homophily, and the dynamics of polarization. *Proceedings of the National Academy of Sciences* **110**, 5791-5796, doi:10.1073/pnas.1217220110 (2013).

48 Lewandowsky, S., Ecker, U. K. H., Seifert, C. M., Schwarz, N. & Cook, J. Misinformation and Its Correction: Continued Influence and Successful Debiasing. *Psychological Science in the Public Interest* **13**, 106-131, doi:10.1177/1529100612451018 (2012).

49 Skinner, B. F. Two Types of Conditioned Reflex and a Pseudo Type. *The Journal of General Psychology* **12**, 66-77, doi:10.1080/00221309.1935.9920088 (1935).

50 Degroot, M. H. Reaching a Consensus. *Journal of the American Statistical Association* **69**, 118-121, doi:10.1080/01621459.1974.10480137 (1974).

51 Ben-Naim, E., Frachebourg, L. & Krapivsky, P. L. Coarsening and persistence in the voter model. *Physical Review E* **53**, 3078-3087, doi:10.1103/PhysRevE.53.3078 (1996).

52 Slanina, F. & Lavicka, H. Analytical results for the Sznajd model of opinion formation. *The European Physical Journal B - Condensed Matter and Complex Systems* **35**, 279-288, doi:10.1140/epjb/e2003-00278-0 (2003).

53 Friedkin, N. E. & Johnsen, E. C. Social influence and opinions. *The Journal of Mathematical Sociology* **15**, 193-206, doi:10.1080/0022250X.1990.9990069 (1990).

54 Lorenz, J. A. N. CONTINUOUS OPINION DYNAMICS UNDER BOUNDED CONFIDENCE: A SURVEY. *International Journal of Modern Physics C* **18**, 1819-1838, doi:10.1142/S0129183107011789 (2007).

55 Lorenz, J. A stabilization theorem for dynamics of continuous opinions. *Physica A: Statistical Mechanics and its Applications* **355**, 217-223, doi:https://doi.org/10.1016/j.physa.2005.02.086 (2005).

56 Wedin, E. & Hegarty, P. A Quadratic Lower Bound for the Convergence Rate in the One-Dimensional Hegselmann–Krause Bounded Confidence Dynamics. *Discrete & Computational Geometry* **53**, 478-486, doi:10.1007/s00454-014-9657-7 (2015).

57 Bhattacharyya, A., Braverman, M., Chazelle, B. & Nguyen, H. L. in *Proceedings of the 4th conference on Innovations in Theoretical Computer Science* 61–66 (Association for Computing Machinery, Berkeley, California, USA, 2013).

58 Hammer, M. R., Bennett, M. J. & Wiseman, R. Measuring intercultural sensitivity: The intercultural development inventory. *International Journal of Intercultural Relations* **27**, 421-443, doi:https://doi.org/10.1016/S0147-1767(03)00032-4 (2003).

59 Holley, R. A. & Liggett, T. M. Ergodic Theorems for Weakly Interacting Infinite Systems and the Voter Model. *The Annals of Probability* **3**, 643-663 (1975).

60 Dittmer, J. C. Consensus formation under bounded confidence. *Nonlinear Analysis: Theory, Methods & Applications* **47**, 4615-4621, doi:https://doi.org/10.1016/S0362-546X(01)00574-0 (2001).

61 Amblard, F., Bouadjio-Boulic, A., Gutiérrez, C. S. & Gaudou, B. in *2015 Winter Simulation Conference (WSC).* 4021-4032.

62 Liu, S., He, L. & Max Shen, Z.-J. On-Time Last-Mile Delivery: Order Assignment with Travel-Time Predictors. *Management Science* **67**, 4095-4119, doi:10.1287/mnsc.2020.3741 (2020).

63 Bien, J. & Tibshirani, R. Hierarchical Clustering With Prototypes via Minimax Linkage. *Journal of the American Statistical Association* **106**, 1075-1084, doi:10.1198/jasa.2011.tm10183 (2011).